\documentclass[preprint,
amsmath,amssymb]{revtex4}
\usepackage{graphicx}

\usepackage{bm}
\usepackage{xy}
\usepackage{epsfig}
\usepackage[usenames]{color}

\usepackage{anysize}

\marginsize{2cm}{2.5cm}{1.5cm}{2cm}

\begin{document}

\title{Parametric oscillators from factorizations employing a constant-shifted Riccati solution of the classical harmonic oscillator}
\author{H.C. Rosu$^1$\footnote{Corresponding author. Phone: +52014448342000; Fax: +52014448342010\\ Electronic mail: hcr@ipicyt.edu.mx} and K.V. Khmelnytskaya$^2$ }
\affiliation{$^1$ IPICyT, Instituto Potosino de Investigacion Cientifica y Tecnologica,\\
Apdo Postal 3-74 Tangamanga, 78231 San Luis Potos\'{\i}, S.L.P., Mexico\\
$^2$ Universidad Autonoma de Queretaro, Centro Universitario, Cerro de las Campanas s/n \\
C.P. 76010 Santiago de Quer\'etaro, Qro., Mexico}

\begin{abstract}
{\footnotesize  \noindent  
We determine the kind of parametric oscillators that are generated in the usual factorization procedure of second-order linear differential equations when one introduces a constant shift of the Riccati solution 
of the classical harmonic oscillator. The mathematical results show that some of these oscillators could be of physical nature. We give the solutions of the obtained second-order differential equations and the values of the shift parameter providing strictly periodic and antiperiodic solutions. We also notice that this simple problem presents parity-time (PT) symmetry. Possible applications are mentioned.}\\




{\footnotesize
}
\end{abstract}

\begin{center} Phys. Lett. A 375(40) 3491-3495, 19 Sept. 2011\\
arXiv: 1008.1441
\end{center}

\maketitle



\medskip

\bigskip

\noindent {\bf 1}.
Riccati nonlinear equations have many applications in physics \cite{nr, r1, r2} and their solutions under the name of superpotentials play an important role in supersymmetric quantum mechanics \cite{susyqm}, which mathematically speaking is the result of a particular case of the factorization algorithm \cite{bo04} as applied to second-order linear differential equations. However, the latter equations form the groundages not only of quantum mechanics but also of many other branches of physics. For example, classical harmonic oscillator equations are amongst the simplest such equations in classical mechanics. Some time ago, factorization techniques have been used by Rosu and Reyes for the damped harmonic classical oscillator \cite{rr98}. In addition, Rosu {\em et al} \cite{rrs98} used a Darboux-embedding technique \cite{kr86} to enclose N underdamped classical modes in a chirp frequency with a hyperbolic secant time profile. The main goal of the present letter is to investigate simple factorization procedures for the free harmonic classical oscillator. As it is well known, particular solutions $R$ of the Riccati equation in its standard form
\begin{equation}  \label{ricco1}
R^{\prime}+ R^2+f=0~, 
\end{equation}
where $f$ is a function of the independent variable, enter the factorization brackets of linear second order differential equations
\begin{equation}
\left( \frac{d}{dt}+R\right) \left( \frac{d}{dt}-R\right)u=0 \equiv u''-(R^2+R')u=0~\equiv u''+fu=0~.
\label{fact1}
\end{equation}%
The connections $R=\frac{u'}{u}$ or $u=e^{\int^t R}$ between the particular solutions of the two equations are also basic results of the factorization method together with the construction of the so-called factorization partner (or Darboux-transformed) equation of Eq.~(\ref{fact1}) obtained by reverting the order of the factorization brackets:
\begin{equation}
\left( \frac{d}{dt}-R\right) \left( \frac{d}{dt }+R\right)v=0\equiv v''-(R^2-R')v=0\equiv v''+(f+2R')v=0~.
\label{fact2}
\end{equation}%
Now, we enquire on the implications of a constant shift of the Riccati solution, i.e.,
$$
R_S(t)=R(t)+S~.
$$
$R_S$ obeys a Riccati equation of the non-standard form:
$$
R^{\prime}_{S}- 2S R_{S}+R^{2}_{S}+(f+S^2)=0~,
$$
which for $S=0$ turns into the standard form (\ref{ricco1}).
The corresponding linear second-order differential equation obtained by substituting $R_S=\frac{u_S'}{u_S}$ is:
$$
u_S''-2Su_S'+(f+S^2)u_S=0
$$
and one can immediately see that a particular solution is $u_S=e^{St}u$. Similarly, $v_S= e^{St}v$ is a particular solution of the supersymmetric equation
$$
v_S''-2Sv_S'+(f+2R'+S^2)v_S=0
$$
related to the non-standard Riccati equation
$$
-R^{\prime}_{S}- 2S R_{S}+R^{2}_{S}+(f+2R'+S^2)=0~.
$$

\noindent However, if we consider the shifted Riccati solution $R_S$ directly in the factorization brackets we obtain the following pair of supersymmetric equations:
\begin{eqnarray}
\Phi''-[(R+S)^2+R']\Phi &=0~,\\
\Psi''-[(R+S)^2-R']\Psi &=0~.
\end{eqnarray}
These equations reduce to the equations for $u$ and $v$ when $S=0$. Nevertheless, their solutions are not connected to the solutions $u$ and $v$ in the same simple way that $u_S$ and $v_S$ are. In the rest of the paper, we will present in full detail the solutions of the latter equations in the case of the classical harmonic oscillator for which $f$ is a constant.\\

\noindent {\bf 2}. The Riccati equation of the classical harmonic oscillator is written as follows:
\begin{equation}  \label{ricco}
R^{\prime}+ R^2+\omega _{0}^2=0~, 
\end{equation}
where $\omega _0$ is the natural frequency of the oscillator. Indeed, using the change of function $R(t)=\frac{u^{'}}{u}$ in Eq.~(1) we get the harmonic oscillator second order differential equation
\begin{equation}  \label{w}
u^{^{\prime\prime}}+\omega_{0}^{2}u=0~, \qquad \omega_{0}^{2}= \mathrm{const}~,
\end{equation}
with the general solution $u=c_1\cos \omega _0 t + c_2\sin \omega _0 t$ and the superposition constants $c_{1}$ and $c_{2}$ fixed through the initial conditions.

\medskip

From the two linearly independent particular solutions $u_1\sim \cos \omega _0 t$ and $u_2 \sim \sin \omega _0 t$,
suppose we choose $u_1$ in the definition of $R$. Then, one gets
$$
R_{u_1}=-\omega _0\tan\omega _0 t~.
$$

\medskip

We consider now the factorization partner equation of Eq.~(\ref{w}), which is given by
\begin{equation}  \label{f1}
\left( \frac{d}{dt}-R_{u_1}\right) \left( \frac{d}{dt }+R_{u_1}\right)v=0~ \longrightarrow v^{^{\prime\prime}} +\omega _{v}^{2}(t)v=0~,
\end{equation}
where
\begin{equation}  \label{geta}
\omega_{v}^{2}(t) =-\omega _{0}^{2}(1+2\tan ^2 \omega _0 t)
\end{equation}
denotes the periodic in time `frequency'  associated through the reverted factorization technique to the constant frequency of the harmonic oscillator.
The linear independent solutions $v_{1}$ and $v_{2}$ are
\begin{equation}\label{linindsol}
v_{1} \sim \frac{\omega _0}{\cos \omega _0 t}~, \qquad  v_{2} \sim \frac{1}{\omega _0\cos \omega _0 t}\left[\frac{\omega _0 t}{2}+\frac{1}{4}\sin 2\omega _0t\right]
~.
\end{equation}

\medskip

We notice that the $u_1$ and $v_1$ oscillator modes fulfill the relationship 
\begin{equation}\label{uv}
u_{1}v_{1}=\omega _0=\mathrm{const}~,
\end{equation}
but not the same happens with the pair ($u_2$,$v_2$) for which $u_2v_2=\int^tu_{1}^{-2}\int^tu_{1}^{2}$ is a function of time. This is a well-known property of second-order linear differential equations related through partner factorizations \cite{Boya98}.\\


\noindent {\bf 3}. We will now investigate the implications of the constant shifting $S$ of the harmonic Riccati solution.
The factorization:
\begin{equation}
\left( \frac{d}{dt}+R_{S}\right) \left( \frac{d}{dt}-R_{S}\right)\mathcal{U}=0~,
\label{calU}
\end{equation}%
leads to:
\begin{equation}
\mathcal{U}^{\prime \prime }+ \Omega _{\mathcal{U}}^{2}(t)
\mathcal{U}=0~,\quad \Omega _{\mathcal{U}}^{2}(t)=\omega _{0}^{2}\left[1-\frac{S^2}{\omega _{0}^{2}}+2\frac{S}{\omega _0}\tan \omega _0t\right]~,   \label{calU bis}
\end{equation}%
whereas the partner factorization 
\begin{equation}
\left( \frac{d}{dt}-R_{S}\right) \left( \frac{d}{dt }+R_{S}\right)
\mathcal{V}=0~,
\label{calV}
\end{equation}%
leads to
\begin{equation}
\mathcal{V}^{\prime \prime }+\Omega _{\mathcal{V}}^{2}(t)\mathcal{V}=0~,
\quad \Omega _{\mathcal{V}}^{2}(t)=-\omega _{0}^{2}\left[1+\frac{S^2}{\omega _{0}^{2}}-2\frac{S}{\omega _0}\tan \omega _0t +2\tan ^{2}\omega _0t\right] ~.  \label{calV bis}
\end{equation}

Equations (\ref{calU bis}) and (\ref{calV bis}) define two classes of oscillator equations describing $\mathcal{U}$ and $\mathcal{V}$ oscillators, respectively. Having time dependent frequency parameters $\Omega _{\mathcal{U, V}}$, these two oscillators are parametric chirps. Moreover, a time periodicity of period $T=\frac{\pi }{\omega_0}$ is embedded therein together with periodic tangent singularities. According to Eqs.~(\ref{calU}) and
(\ref{calV}) these two chirps are connected by the same Riccati solution given by:
\begin{equation}
R_{S}=-\omega _0\tan \omega _0t+S~.  \label{reg1}
\end{equation}%
Obviously, when $S\rightarrow 0$ we have $R_S\rightarrow R_{u_{1}}$ and the oscillators $\mathcal{U}$ and $\mathcal{V}$ are very similar to the unshifted oscillators $u$ and $v$ in this limit.\\

The linear independent solutions $\mathcal{U}_1$ and $\mathcal{U}_2$ have the following form
\begin{equation}
\mathcal{U}_{1}(t)\sim e^{-i\Omega _{S}t}\,{}_{2}F_{1}\left( 1,-\frac{iS}{\omega _0};2-\frac{iS}{\omega _0};-e^{-2i\omega _0t}\right) ~,
\label{2-fcfin1}
\end{equation}
\begin{equation}
\mathcal{U}_{2}(t)\sim
e^{i\Omega _{S}t}{}_{2}F_{1}\left(\frac{iS}{\omega _0},-1;\frac{iS}{\omega _0};-e^{-2i\omega _0t}\right)=
e^{i\Omega _{S}t}(2\cos ^2\omega _0 t-i \sin 2\omega _0 t)\sim e^{St}\cos \omega _0 t ~,  \label{2-fcfin2}
\end{equation}%
where $\Omega _{S}=(1-\frac{iS}{\omega _0})\omega _0$ and $_{2}F_{1}\left( a,b;c;z\right)$ is the hypergeometric series. Unlike $\mathcal{U}_{2}(t)$, the hypergeometric function in $\mathcal{U}_{1}(t)$ cannot be reduced to elementary functions.
The simple convergence condition $\Re (a+b-c)<0$ for the hypergeometric series is fulfilled for these solutions for all real values of $t$.
We discuss more about the mode $\mathcal{U}_{1}(t)$ in the Appendix. Moreover, one can notice immediately that the solutions (\ref{2-fcfin1}),(\ref{2-fcfin2}) 
have a convenient Floquet-Bloch form. 
In addition, the parameter $S$ affects only the period of the phases $e^{\pm i\Omega _{S}t}$ of the solutions but not that of their periodic part.\\

On the other hand, the linear independent solutions $\mathcal{V}$ are given by:
\begin{equation}
\mathcal{V}_{1}(t)\sim\frac{e^{-St}}{\cos \omega _0t}~,
\label{g1}
\end{equation}
\begin{equation}
\mathcal{V}_{2}(t) \sim e^{St}\left(\frac{\omega _{0}^{2}}{\cos\omega _0t}+2S\omega _0\sin \omega _0 t+2S^2\cos \omega _0t \right)
~.  \label{g2}
\end{equation}

Both solutions $\mathcal{V}_{1}(t)$ and $\mathcal{V}_{2}(t)$ are singular at $t =\pm \frac{\pi}{2\omega _0}$ and at their odd multiples because of the presence of the cosine function in the denominator. Again, one can find a pair of solutions ($\mathcal{U}$, $\mathcal{V}$) for which  $\mathcal{U} \mathcal{V}=$ const, in this case ($\mathcal{U}_2$, $\mathcal{V}_1$).\

The solutions (\ref{2-fcfin1}),(\ref{2-fcfin2}) and the exponential factor  $e^{St}$ in (\ref{g1}),(\ref{g2}) are bounded if and only if the ``quasifrequency" $\Omega _{S}$ has a real value, or equivalently
\begin{equation}
\omega _0-iS\in \mathbf{R}~.  \label{cond-1}
\end{equation}%
Taking into account that $\omega _0\in \mathbf{R}$, the last condition reads as $S=is,\,s\in \mathbf{R}$. Notice that for a purely imaginary shift parameter, the chirp frequencies $\Omega ^{2} _{\mathcal{U}}$ and $\Omega^{2} _{\mathcal{V}}$ are related through $\Omega^{2} _{\mathcal{U},\mathcal{V}}(-t)=\Omega^{2*} _{\mathcal{U},\mathcal{V}}(t)$, where * denotes the complex conjugation operation. Such a property is called parity-time (PT) symmetry in quantum mechanics, where it has been thoroughly investigated since it has been found that the spectra of the PT-symmetric operators are real \cite{bender07}.
Additionally, by inspecting the solutions (\ref{2-fcfin1}),(\ref{2-fcfin2}) and (\ref{g1}),(\ref{g2}) we note that they are periodic for $s=(2m-1)\omega _0$, $m=0,\pm 1,\pm 2,...$\ and antiperiodic for $s=2m\omega _0$, $m=\pm 1,\pm 2,...~$.

Plots of the $S=5i$ periodic and $S=6i$ antiperiodic solutions are presented in Figures 1,2 and 3,4, respectively, all of them for $\omega _0=1$. The $\mathcal{U}$ modes are regular in both their real and imaginary parts, while the periodic and antiperiodic $\mathcal{V}$ modes have periodic singularities in their imaginary parts and real parts, respectively.

For imaginary $S$, the frequency parameters are:
\begin{equation}
\Omega _{\mathcal{U}}^{2}(t)=\omega _{0}^{2}\left[1+\frac{s^2}{\omega _{0}^{2}}+2i\frac{s}{\omega _0}\tan \omega _0t\right]~
\label{calU bis1}
\end{equation}
and
\begin{equation}
\Omega _{\mathcal{V}}^{2}(t)=\omega _{0}^{2}\left[-1+\frac{s^2}{\omega _{0}^{2}}-2\tan ^{2}\omega _0t+2i\frac{s}{\omega _0}\tan \omega _0t\right] ~.  \label{calV bis1}
\end{equation}
 Equation (\ref{calU bis1}) shows that the time-dependent (chirp) feature is present only in the imaginary part of the frequency parameter for the $\mathcal{U}$ oscillators, whereas their real part looks very common. On the other hand, the $\mathcal{V}$ oscillators are chirpy in both their real and imaginary parts of their frequency parameters and their real part could be even negative.

\begin{figure}[ptb]
\centering
\includegraphics[height=4.4622in,width=5.9055in] {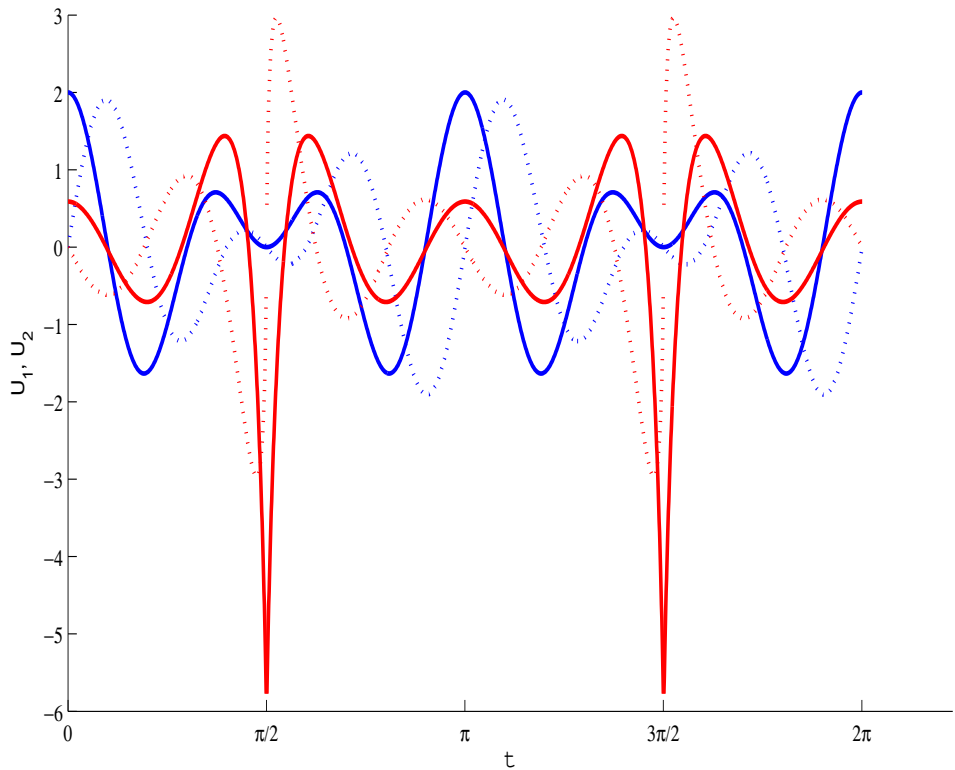}
\caption{The real (solid lines) and imaginary (dotted lines) parts of the
periodic solutions $\mathcal{U}_{1}(t)$ (in red) and $\mathcal{U}_{2}%
(t)$ (in blue) for the shift parameter $S=5i$.}%
\label{figureS1}%
\end{figure}

\begin{figure}[ptb]
\centering
\includegraphics[height=4.4622in,width=5.9055in] {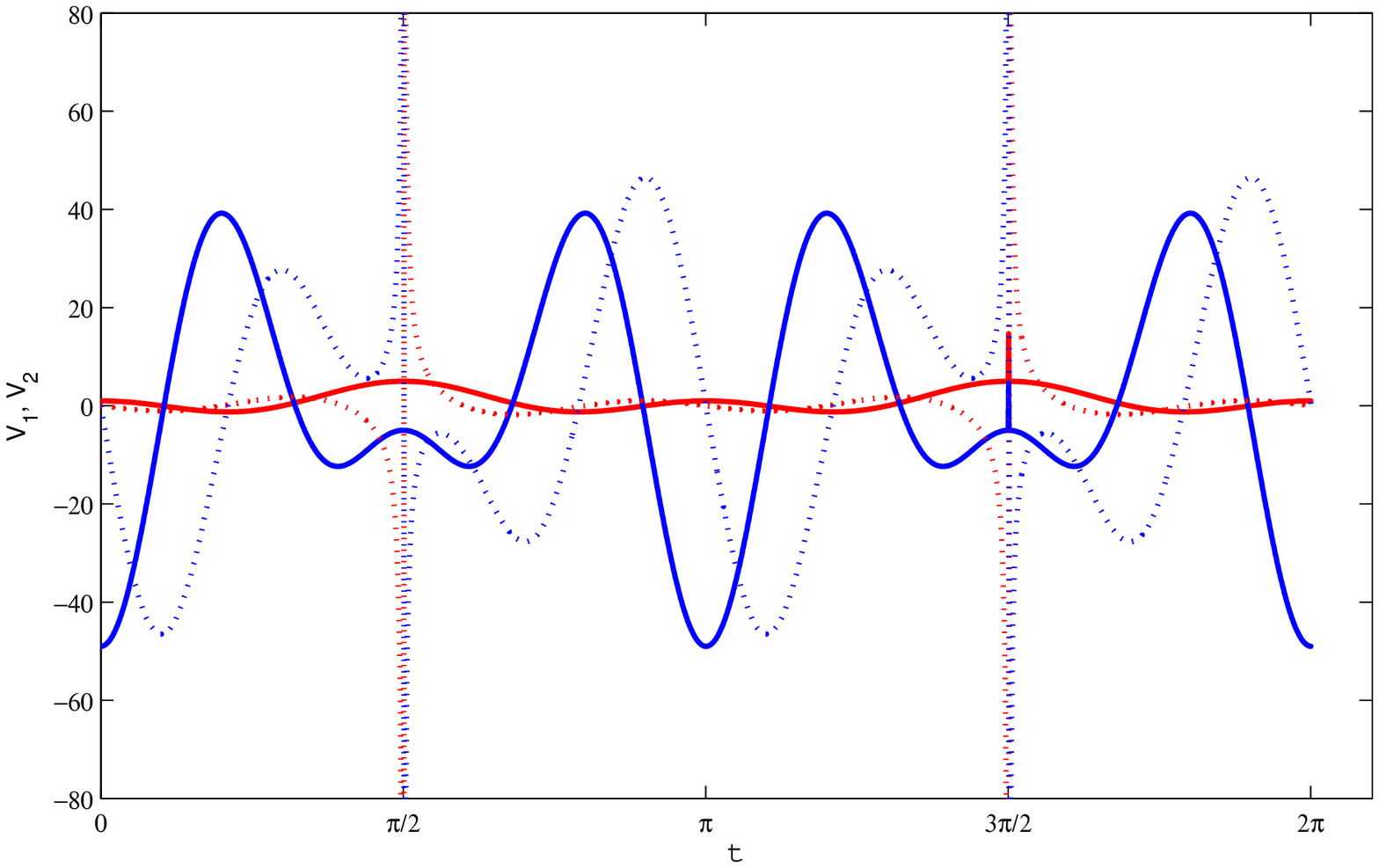}
\caption{The real (solid line) and imaginary (dotted lines) parts of the
periodic solutions $\mathcal{V}_{1}(t)$ (in red) and $\mathcal{V}_{2}(t)$ (in blue) for the shift parameter $S=5i$.}
\label{figureS2}%
\end{figure}

\begin{figure}[ptb]
\centering
\includegraphics[height=4.4622in,width=5.9055in] {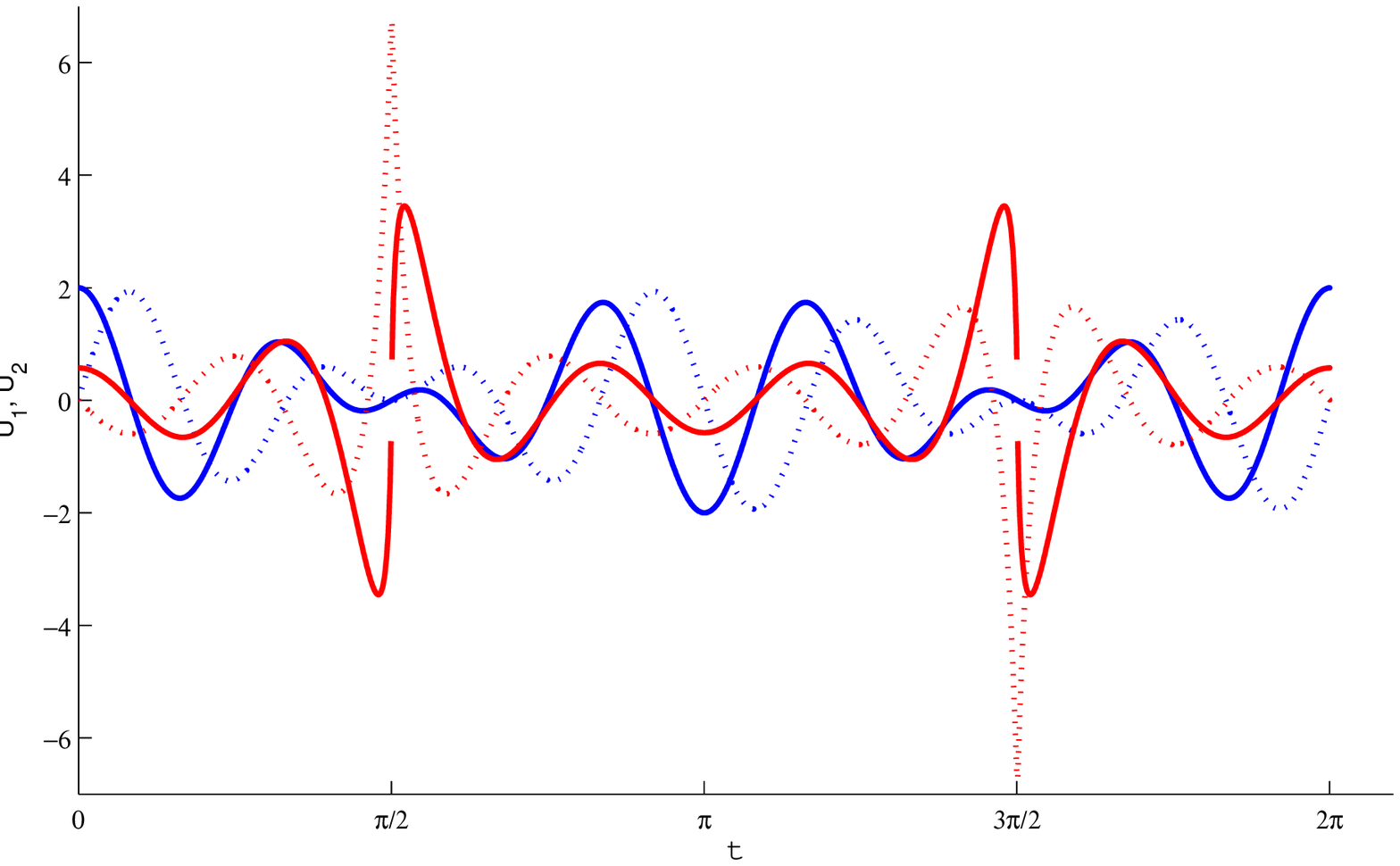}
\caption{The real (solid lines) and imaginary (dotted lines) parts of the
antiperiodic solutions $\mathcal{U}_{1}(t)$ (in red) and $\mathcal{U}_{2}(t)$ (in blue) for the shift parameter $S=6i$.}%
\label{figureS3}%
\end{figure}

\begin{figure}[ptb]
\centering
\includegraphics[height=4.4622in,width=5.9055in] {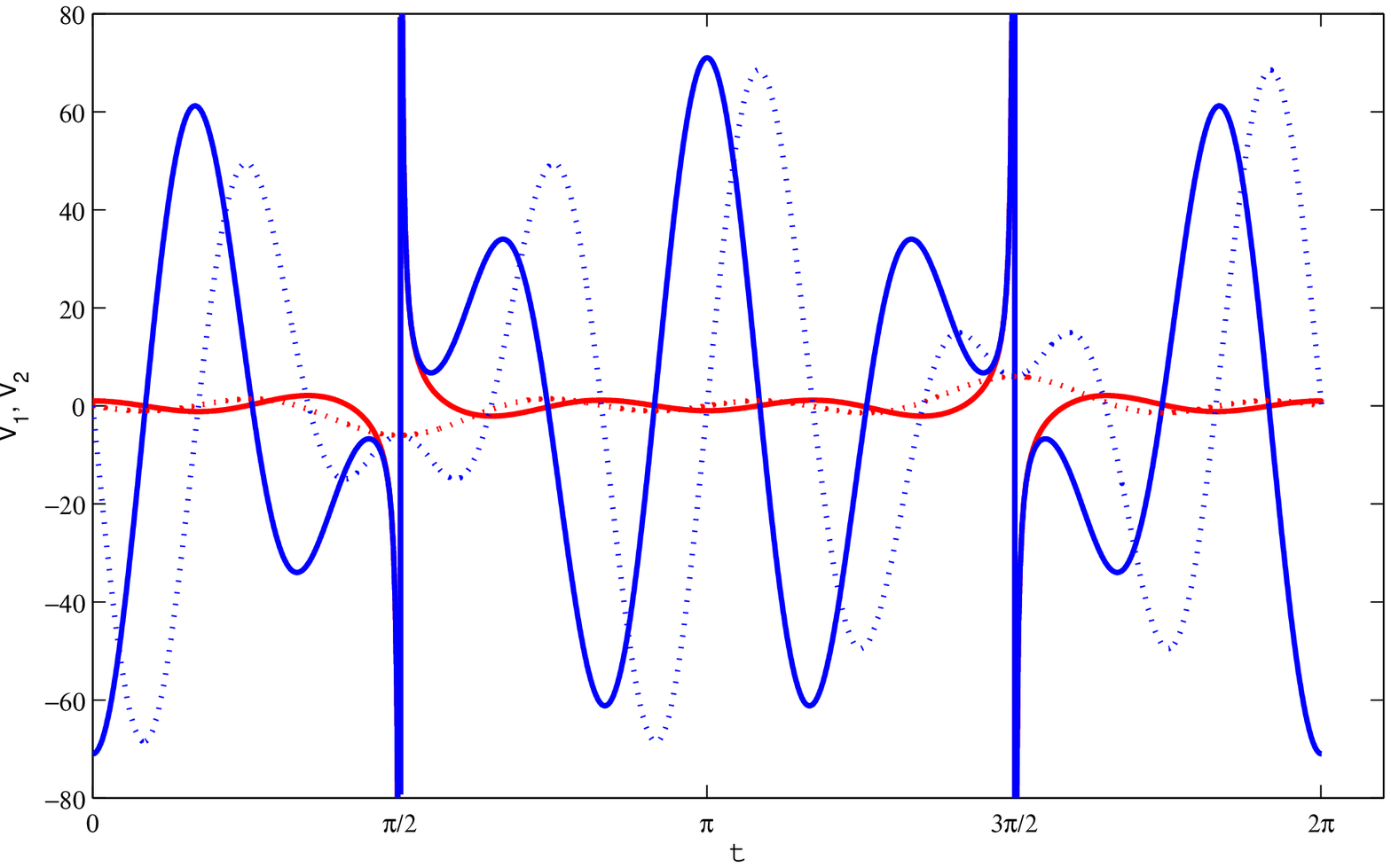}
\caption{The real (solid lines) and imaginary (dotted lines) parts of the
antiperiodic solutions $\mathcal{V}_{1}(t)$ (in red) and $\mathcal{V}_{2}(t)$ (in blue) for the shift parameter $S=6i$.}
\label{figureS4}%
\end{figure}


\medskip
\medskip


\noindent {\bf 3}. In summary, in this Letter, we introduced a class of chirp oscillators that are related to the classical harmonic oscillators through a constant shift of a Riccati particular solution of the harmonic oscillator. Non-trivial results are obtained when the shift parameter is purely imaginary and in particular we provide the values of the parameter when the oscillatory modes are strictly periodic or antiperiodic.

We also found that this simple problem displays PT symmetry.
Simple classical problems with PT symmetry have been also discussed by Bender \cite{bender07}. However, the way we introduce the PT symmetry is by a constant imaginary shift of a particular Riccati solution of the problem which is different of the standard procedure in which one employs anharmonic imaginary contributions in the potential function.

One may ask in what particular context these shifted oscillators could be found. Indeed, it is not easy to think of a direct physical occurrence of the Riccati solutions. However, we have found recently that in barotropic Friedman-Robertson-Walker cosmologies the Hubble parameter is just the solution of a special type of Riccati equation, which is the same as that of a classical harmonic oscillator unless a linear expression of the adiabatic index of the cosmological fluid substitutes the natural frequency $\omega _0$ of the harmonic oscillator and instead of the Newtonian time one works in the conformal time \cite{hak}. Another more terrestrial application could be in the context of parametric excitation which results from a periodic variation in an energy storage element in a system. To see this, it is sufficient to write the shifted oscillator equations for imaginary $S$ in the form \cite{wiki-po}
\begin{equation}\label{p-exc1}
\frac{d^2\mathcal{U}}{dt^2}+\omega_{0}^{2}\mathcal{U}=-\omega_{0}^{2}h(t)\mathcal{U}~, \qquad
\frac{d^2\mathcal{V}}{dt^2}+s^{2}\mathcal{V}=-s^{2}g(t)\mathcal{V}~,
\end{equation}
where $h(t)=2i\sigma \tan \omega _0t-\sigma^2$ and
$g(t)=2i\sigma^{-1}\tan \omega _0t-2\sigma^{-2}\tan^2\omega _0t-\sigma^{-2}$, ($\sigma=\frac{s}{\omega _0}$), respectively.
The equations in (\ref{p-exc1}) represent the parametric excitation of simple harmonic oscillators (or, bandpass filters), of natural frequencies $\omega _0$ and $s$,
driven by the signals $-\omega_{0}^{2}h(t)\mathcal{U}$ and $-s^{2}g(t)\mathcal{V}$ proportional to their responses $\mathcal{U}$ and $\mathcal{V}$, respectively. Thus, parametric excitation experiments with the particular pumpings $h$ and $g$ are a way of studying these types of shifted oscillators and their correlations.

\newpage

{\bf Appendix}: Remarks on the solution $\mathcal{U}_{1}(t)=e^{-i\Omega _{S}t}\,{}_{2}F_{1}\left( 1,-\frac{iS}{\omega_0};2-\frac{iS}{\omega _0};-e^{-2i\omega _0t}\right)$\\

Formula 15.3.4 in Abramowitz and Stegun \cite{as} (in Linear transformation formulas, page 559) reads:

\begin{equation}\label{as}
F(a,b;c;z)=(1-z)^{-a}F\left(a,c-b;c;\frac{z}{z-1}\right)~.
\end{equation}

Using Eq.~(\ref{as}) for the HGM series of the $\mathcal{U}_{1}(t)$ we get:
\begin{equation}
{}_{2}F_{1}\left( 1,-\frac{iS%
}{\omega _0};2-\frac{iS}{\omega _0};-e^{-2i\omega _0 t}\right)=
\frac{e^{i\omega _0t}}{2\cos\omega _0t}\,{}_{2}F_{1}\left(1,2;2-\frac{iS}{\omega _0};\frac{1}{1+e^{2i\omega _0t}}\right)~.
\end{equation}
Thus:
\begin{equation}
\mathcal{U}_{1}(t)=\frac{1}{2}\frac{1}{e^{St}\cos \omega _0t} \,{}_{2}F_{1}\left(1,2;2-\frac{iS}{\omega _0};\frac{1}{1+e^{2i\omega _0t}}\right)~.
\end{equation}
On the other hand, we can use the following relationship:
\begin{equation}
(1-z)^{-p}={}_{2}F_{1}\left(p,b;b;z\right)
\end{equation}
for $p=1$ and $S=0$ in (11), which leads to:
\begin{equation}
\mathcal{U}_{1}(t)=\frac{1}{2}\frac{1}{\cos\omega _0t} 2\cos \omega _0t \, e^{-i\omega _0t}\,=e^{-i\omega _0t}
\end{equation}
as expected.\\

However, if $S\neq 0$, we get the complicated HGM formula:
\begin{equation}
\mathcal{U}_{1}(\eta )=\frac{1}{2}\frac{1}{\cos\omega _0t}\Gamma\left(2-\frac{iS}{\omega _0}\right)
\sum _{n=0}^{\infty}
\frac{(n+1)!}{\Gamma(n+2-\frac{iS}{\omega _0})}(1+e^{2i\omega _0t})^{-n}~.
\end{equation}

\end{document}